\documentclass[]{pasj01}
\Received{2022/12/16}
\usepackage{color}
\usepackage[separate-uncertainty, angle-symbol-over-decimal]{siunitx}
\usepackage{url}
\usepackage{bm}
\DeclareAbbreviation\icarus{Icarus}
\DeclareAbbreviation\psj{PSJ}

\begin{document} 
\title{Simultaneous Multicolor Photometry of the DESTINY$^{+}$ target asteroid (3200) Phaethon}

%%% Authors ===================================================================
%%% Core members (3)
\author{Jin \textsc{Beniyama}\altaffilmark{1,2,*}}
\email{beniyama@ioa.s.u-tokyo.ac.jp}
\author{Tomohiko \textsc{Sekiguchi}\altaffilmark{3}}
\author{Daisuke \textsc{Kuroda}\altaffilmark{4,5}}
%%%% DESTINY^{+} members (6->4)
\author{Tomoko \textsc{Arai}\altaffilmark{6}}
\author{Ko \textsc{Ishibashi}\altaffilmark{6}}
\author{Masateru \textsc{Ishiguro}\altaffilmark{7,8}}
\author{Fumi \textsc{Yoshida}\altaffilmark{6,9}}
\author{Hiroaki \textsc{Senshu}\altaffilmark{6}}
\author{Takafumi \textsc{Ootsubo}\altaffilmark{10}}
%%%% IoA members (3) (Ohsawa-san moved to NAOJ in August, 2022) 
\author{Shigeyuki \textsc{Sako}\altaffilmark{1,11,12}}
\author{Ryou \textsc{Ohsawa}\altaffilmark{10}}
\author{Satoshi \textsc{Takita}\altaffilmark{1}}
%%%% SNU members (2)
\author{Jooyeon \textsc{Geem}\altaffilmark{7,8}}
\author{Yoonsoo P. \textsc{Bach}\altaffilmark{7,8}}
%%% End Authors ===============================================================

%%% Affiliations ==============================================================
\altaffiltext{1}{%
Institute of Astronomy, Graduate School of Science,
The University of Tokyo, 2-21-1 Osawa, Mitaka, Tokyo 181-0015, Japan}
\altaffiltext{2}{%
Department of Astronomy, Graduate School of Science,
The University of Tokyo, 7-3-1 Hongo, Bunkyo-ku, Tokyo 113-0033, Japan}
\altaffiltext{3}{%
Asahikawa Campus, Hokkaido University of Education, 
Hokumon, Asahikawa, Hokkaido 070-8621, Japan}
\altaffiltext{4}{%
Okayama Observatory, Kyoto University, 
3037-5 Honjo, Kamogata-cho, Asakuchi, Okayama 719-0232, Japan}
\altaffiltext{5}{%
Bisei Spaceguard Center, Japan Spaceguard Association, 
1716-3 Okura, Bisei, Ibara, Okayama 714-1411, Japan}
\altaffiltext{6}{%
Planetary Exploration Research Center,
Chiba Institute of Technology, 2–17–1 Tsudanuma, Narashino,
Chiba, 275–0016, Japan}
\altaffiltext{7}{%
Department of Physics and Astronomy, 
Seoul National University, 1 Gwanak-ro, Gwanak-gu, 
Seoul 08826, Republic of Korea}
\altaffiltext{8}{%
SNU Astronomy Research Center, 
Seoul National University, 
1 Gwanak-ro, Gwanak-gu, Seoul 08826, Republic of Korea}
\altaffiltext{9}{%
School of Medicine, Department of Basic Sciences,
University of Occupational and Environmental Health, 1-1 Iseigaoka,
Yahata, Kitakyusyu 807-8555, Japan}
\altaffiltext{10}{%
National Astronomical Observatory of Japan, 
2-21-1 Osawa, Mitaka, Tokyo 181-8588, Japan}
\altaffiltext{11}{%
UTokyo Organization for Planetary Space Science, 
The University of Tokyo, 7-3-1 Hongo, Bunkyo-ku, 
Tokyo 113-0033, Japan}
\altaffiltext{12}{%
Collaborative Research Organization for
Space Science and Technology, 
The University of Tokyo, 7-3-1 Hongo, 
Bunkyo-ku, Tokyo 113-0033, Japan}
%%% End Affiliations ==========================================================

%%% List of Key Words: https://academic.oup.com/pasj/pages/Pasj_Keywords 

\maketitle

%%% Abstract less than 300 words (180/300)
\begin{abstract}
Accurate estimation of brightness of (3200) Phaethon up to lower phase angles 
are essential for planning of the on-board camera of the DESTINY$^{+}$ mission.
We have carried out intensive observations of Phaethon in the optical wavelength 
($g$, $r$, and $i$)
with the TriCCS camera on the Seimei 3.8\,m telescope in October and November, 2021. 
We derived the absolute magnitude $H_\mathrm{V}$ and the slope parameter $G$ of Phaethon
as $H_\mathrm{V}=14.23\pm0.02$ and $G=0.040\pm0.008$ 
from multiple photometric observations including lower phase angles down to $\sim$9$^{\circ}$
with the $H$-$G$ model.
Using the $H_\mathrm{V}$ value and the geometric albedo of Phaethon derived in 
previous polarimetric studies, we estimated that the Phaethon's 
diameter is within a range of 5.22 to 6.74\,km,
which is consistent with radar and occultation observations.
With the linear model, we derived $H_\mathrm{V}=14.65\pm0.02$,
which corresponds to a diameter range of 4.30 to 5.56\,km.
Our simultaneous tricolor lightcurves of Phaethon indicate that 
no rotational spectral variations larger than 0.018 and 0.020 mag in the g-r and r-i colors, 
possibly related to inhomogeneity of the surface material and/or structure,
are seen at the 2021 apparition.
\end{abstract}

%%% Add main parts
\section{Introduction}
% About (3200) Phaethon
The asteroid (3200) Phaethon is one of the most attractive asteroids: 
dust activity near the perihelion 
(Jewitt \& Li\,\yearcite{Jewitt2010}; \cite{Jewitt2013a}; Li \& Jewitt\,\yearcite{Li2013}; Hui \& Li\,\yearcite{Hui2017}),
near-Sun and near-Earth orbits \citep{Ohtsuka2009},
and the parent body of the Geminid meteor shower (Williams \& Wu \yearcite{Williams1993}).
Phaethon was discovered in images of the Infrared Astronomy Satellite (IRAS) 
in 1983 (Green \& Kowal\,\yearcite{Green1983}).
%% The same as \textit{Phaethon} in the Greek Mythology,
Phaethon approaches Sun every about 1.4 years
with a large eccentricity, $\sim$0.89, and an extremely small perihelion distance, $\sim$0.14 au.
The perihelion activities were reported in 2009, 2012, and 2016 
by observations with the Solar TErrestrial RElations Observatory (STEREO) spacecraft
(Jewitt \& Li\,\yearcite{Jewitt2010}; \cite{Jewitt2013a}; Li \& Jewitt\,\yearcite{Li2013}; Hui \& Li\,\yearcite{Hui2017}).
Phaethon is taxonomically classified as 
B-type in the Bus taxonomy (\cite{Bus1999}; Bus \& Binzel\,\yearcite{Bus2002}).
%% F-type in Tholen taxonomy \citep{Tholen1984}? Still unclear?

Phaethon is selected as a target of the 
exploration mission Demonstration and Experiment of Space Technology for INterplanetary voYage 
Phaethon fLyby and dUst Science, DESTINY$^{+}$ \citep{Arai2018}.
The DESTINY$^{+}$ is a mission proposed for the JAXA/ISAS Epsilon class small program
and will be launched in 2024 \citep{Arai2018, Arai2022, Ozaki2022}.
Since the rendezvous is difficult for Phaethon with a large relative velocity 
arise from the large eccentricity and orbital inclination, 
the DESTINY$^{+}$ will flyby Phaethon in 2028 \citep{Arai2018}.
High-resolution imaging 
with the panchromatic telescopic camera (TCAP) and the VIS-NIR multiband camera (MCAP)
will be performed at the flyby.
% Purposes
% Porblem1 size, H,G
% Distingish "surface brightness" and "brightness".
To succeed just one DESTINY$^{+}$ flyby with Phaethon,
we need to know its surface brightness with high accuracy beforehand.
It is desired that the on-board camera setting, such as sensor gain and exposure time, 
are fixed before the flyby since the time is limited to change the camera 
setting just before the approach due to the large relative velocity at the flyby.

The disk-integrated brightness dependence on the solar phase angle is called a phase curve.
The phase curve provides an estimate of an absolute magnitude $H$, 
the apparent magnitude of an object located 1 au apart from Sun and Earth at zero phase angle.
In general, we inevitably derive the $H$
by extrapolating the phase curve 
because it is difficult to obtain photometric data at the zero phase angle.
For Phaethon, \citet{Ansdell2014} reported the absolute magnitude in the Johnson $R$-band, 
$H_\mathrm{R}$, as 13.90 in observations at phase angles of 12 to 83$^{\circ}$.
\citet{Tabeshian2019} derived the $H_\mathrm{R}$ as $13.28\pm0.02$
in observations at phase angles of 20 to 100$^{\circ}$.
The derived absolute magnitudes have a discrepancy.
In addition, there may be a systematic uncertainty in the absolute magnitude of Phaethon
since no observation at phase angles $\alpha$ below $12^{\circ}$ has been reported. 
That is to say, the disk-integrated brightness for Phaethon is still unclear, especially at lower phase angles.
To derive the absolute magnitude with high accuracy, 
additional observations at lower phase angles are necessary.
% Problem2 inhomogeneity
An investigation of the surface inhomogeneity of Phaethon is also one of the 
most important tasks to do before the DESTINY$^{+}$ flyby.
Several surface features on Phaethon such as concavities and boulders
have been reported by radar observations \citep{Taylor2019}.
To know the local terrain of Phaethon's surface by ground-based observations 
before the flyby is important for the DESTINY$^{+}$ mission
because they can adjust the flyby timing to observe the interesting side of Phaethon.

Rotational color variations can reflect the surface inhomogeneity of asteroids.
In previous studies, color variations of Phaethon
were detected by spectroscopic
\citep{Licandro2007, Kareta2018, Lazzarin2019, Ohtsuka2020}
and multicolor photometric observations \citep{Tabeshian2019, Lin2020}.
Recently, \citet{MacLennan2022b} found an evidence of the heterogeneity 
of surface grain size on Phaethon among different latitudes
with thermophysical modeling (TPM) using the latest shape model 
constructed with radar data as well as optical lightcurves.
However, uniform colors of Phaethon in longitude are also reported \citep{Borisov2018, Lee2019}. 
The inhomogeneity of Phaethon's surface is still debated and follow-up observations in other apparitions are required.

% Purposes and Paper flow
The aims of this study are to constrain $H_\mathrm{V}$ and $G$ parameters 
and to investigate Phaethon's surface inhomogeneity before the DESTINY$^{+}$ flyby.
We carried out simultaneous multicolor photometry of Phaethon in 2021.
Simultaneous multicolor observations allow us to achieve accurate measurements of 
the surface colors since the simultaneity cancels out the atmospheric variations in the measurements.
In this paper, we will describe observations and data reduction in section 2.
The results are compared with previous studies in section 3.
In section 4, we discuss the absolute magnitude, diameter, and inhomogeneity of Phaethon.

\section{Observations and data reduction}

\subsection{Observations}
We conducted multicolor photometry with the TriColor CMOS Camera 
and Spectrograph (TriCCS) on the Seimei 3.8\,m telescope \citep{Kurita2020}
in Okayama, Japan.
The observing conditions are summarized in table \ref{tab:obs}.
\begin{table*}
           \tbl{Summary of the observations.\footnotemark[$*$]}{%
           % 8 columns
           \begin{tabular}{cccccccccc}
             \hline
             Obs. Date  & $T_{\mathrm{exp}}$ & $N_{\mathrm{img}}$ & $V$   & Velocity                           & $\alpha$   & $\Delta$  & $r_\mathrm{h}$  & Airmass & Note  \\ 
             (UTC)      & (s)                &                    & (mag) &  ($\mathrm{arcsec\,\,h^{-1}}$) & ($^\circ$) & (au)      & (au) &    &   \\
             \hline
             2021-10-27 14:03:30--20:14:54&30 & 585 & 17.9 &46.6 & 15.9 &1.4602 & 2.3133 & 1.01--1.28 &  \\
2021-10-28 11:56:47--20:08:57&60 & 408 & 17.9 &47.4 & 15.6 &1.4498 & 2.3107 & 1.01--1.87 & photometric\\
2021-10-29 13:29:30--20:19:49&60 & 318 & 17.8 &49.2 & 15.2 &1.4380 & 2.3075 & 1.01--1.34 &  \\
2021-11-10 10:34:33--20:27:50&60 & 304 & 17.5 &64.8 & 10.3 &1.3286 & 2.2683 & 1.00--1.89 & photometric\\
2021-11-11 10:27:49--20:13:04&60 & 357 & 17.4 &66.0 & 10.0 &1.3215 & 2.2647 & 1.00--1.87 &  \\
2021-11-12 10:19:30--20:25:47&60 & 435 & 17.4 &67.2 & 9.7 &1.3148 & 2.2611 & 1.00--2.01 & photometric\\
2021-11-24 12:40:26--18:56:40&60 & 186 & 17.2 &79.1 & 9.1 &1.2614 & 2.2126 & 1.00--1.99 &  \\
2021-11-25 09:30:53--17:15:42&60 & 267 & 17.2 &78.7 & 9.3 &1.2597 & 2.2088 & 1.00--1.60 & photometric\\
2021-11-26 09:31:57--18:48:41&60 & 408 & 17.2 &79.1 & 9.6 &1.2580 & 2.2044 & 1.00--2.06 & photometric\\
\hline
            \end{tabular}
            }\label{tab:obs}
            \begin{tabnote}
              \footnotemark[$*$] 
              Observation time in UTC (Obs. Date), exposure time ($T_{\mathrm{exp}}$),
              and the number of images ($N_\mathrm{img}$) for each night are listed.
              Predicted $V$-band apparent magnitude ($V$), 
              angular rate of change in apparent RA and DEC (Velocity),
              phase angle ($\alpha$),
              distance between Phaethon and observer ($\Delta$),
              and 
              distance between 
              Phaethon and Sun ($r_\mathrm{h}$) at the observation starting time
              are from NASA JPL/HORIZONS
               as of 2022-11-8 (UTC).
              Elevations to calculate air mass range (Airmass) are 
              also from NASA JPL/HORIZONS.
            \end{tabnote}\end{table*}
We obtained $g$, $r$, and $i$-band images with the Pan-STARRS filters 
simultaneously with TriCCS.
The field of view of each 2k $\times$ 1k CMOS image sensor is $\ang{;12.6;}\times\ang{;7.5;}$ with a pixel scale of 0.350 arcsec.

Phaethon was observed on nine nights in October and November, 2021.
On the first day, October 27th, we set the exposure time as 30\,s 
and realized that the signal-to-noise ratio was not enough to discuss the surface inhomogeneity.
Then, we set nominal exposure times as 60\,s from October 28th to achieve a 
high signal-to-noise ratio.

\begin{figure}
\includegraphics[width=80mm]{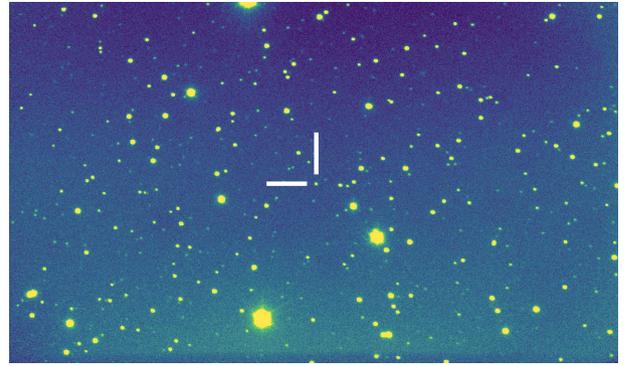}
\caption{%
   Image of (3200) Phaethon in $r$-band with a 60 s exposure in October 28th, 2021
   taken by TriCCS.
   Field of view covers $\ang{;12.6;}\times\ang{;7.5;}$.
   North is to the top and East is to the left.}
\label{fig:image}
\end{figure}

The telescope was operated in the non-sidereal tracking mode to follow Phaethon 
during our observations with ephemerides 
from NASA JPL/HORIZONS\footnote{https://ssd.jpl.nasa.gov/?horizons}.
Elongations of the point spread functions (PSF) of 
stars in the images (hereinafter referred to as reference stars)
were negligible
since Phaethon's angular velocities in the observations 
were small enough compared to a typical PSF, 2--5 arcsec 
in full width at half maximum (FWHM).
The observation fields were crowded with reference stars
since Phaethon was located close to the galactic plane at the time of our observations (galactic latitude $|b| < 20^{\circ}$, figure \ref{fig:image}).

\subsection{Data reduction}

\subsubsection{Photometry}
After standard corrections (bias subtraction, dark subtraction, and flat-field correction),
cosmic ray events were removed using the \texttt{Python} package \texttt{astroscrappy}
\citep{McCully2018}
based on Pieter van Dokkum's \texttt{L.A.Cosmic} algorithm
\citep{vanDokkum2001}.
The standard circular aperture photometry was done on Phaethon and reference stars
in each frame using the SExtractor-based \texttt{Python} package \texttt{sep}
(Bertin \& Arnouts\,\yearcite{Bertin1996};\,\cite{Barbary2015}).
The aperture radius was set to 2 times larger than the FWHM
of the PSF of reference stars.
The FWHM of the PSF in each frame was calculated using reference stars before the photometry.
The light-travel time correction was done to obtain the time-series colors and magnitudes (Harris \& Lupishko\,\yearcite{Harris1989}).

Stars meet any criterion below were not used in the analysis:
uncertainties in $g$, $r$, or $i$-band magnitudes 
in Pan-STARRS Data Release 2 (DR2, \cite{Chambers2016}) are larger than 0.05 mag,
$(g-r)_\mathrm{PS}> 1.1$, 
$(g-r)_\mathrm{PS}< 0.0$, 
$(r-i)_\mathrm{PS}> 0.8$,
or 
$(r-i)_\mathrm{PS}< 0.0$, where
$(g - r)_{\mathrm{PS}}$ and $(r - i)_{\mathrm{PS}}$ are colors in Pan-STARRS system.
Also, we removed photometric results 
close to the edges of the image (200 pixels from the edge) or with any other sources in its aperture frame by frame.
Extended sources, possible quasars, and variable stars were removed 
using the \texttt{objinfoflag} and \texttt{objfilterflag} in Pan-STARRS DR2.

\subsubsection{Color and magnitude derivation}
Simultaneous multicolor observations are a reliable method to
measure the accurate colors of asteroids.
We derived the Phaethon's surface colors 
in the slightly different manner in \citet{Jackson2021a}\footnote{The differences are filter bands and airmass dependences of first items in equations (\ref{eq:g_CT})--(\ref{eq:i_CT}).
While our first items, instrumental magnitudes, depend on airmass, theirs do not.}.
We used the linear relationship between instrumental colors and colors in the Pan-STARRS system as follows:
\begin{eqnarray}
    (g - r)_{\mathrm{PS}, m}^n &= \mathrm{CTG}_{g-r} (g - r)_{\mathrm{inst}, m}^n  + \mathrm{CTI}_{g-r}^n, \label{eq:g-r_CTG} \\
    (r - i)_{\mathrm{PS}, m}^n &= \mathrm{CTG}_{r-i} (r - i)_{\mathrm{inst}, m}^n  + \mathrm{CTI}_{r-i}^n, \label{eq:r-i_CTG}
\end{eqnarray}
where $m$ is the index of the object, 
$n$ is the index of the frame,
$(g - r)_{\mathrm{inst}, m}^n$ and $(r - i)_{\mathrm{inst}, m}^n$ are
instrumental colors of $m$-th object on the $n$-th frame, 
$\mathrm{CTG}_{g-r}$ and $\mathrm{CTG}_{r-i}$ are 
color transformation gradients (CTGs) of the $g-r$ and $r-i$ colors, respectively,
and 
$\mathrm{CTI}_{g-r}^n$ and $\mathrm{CTI}_{r-i}^n$ are color transformation intercepts (CTIs)
of the $g-r$ and $r-i$ colors on the $n$-th frame, respectively.
First, we derived the CTGs at the night 
by the linear fitting of the photometric results of the reference stars during the night.
To determine the unique CTGs at the night,
the photometric results were shifted frame by frame
to cancel out the atmospheric variations in CTIs.
After the derivation of CTGs, 
the CTIs in each frame were calculated with the derived CTGs
with equations (\ref{eq:g-r_CTG}) and (\ref{eq:r-i_CTG}).
Phaethon's colors were derived from the instrumental colors, fixed CTGs, and CTIs.
We computed the propagated uncertainties of 
$(g-r)_{\mathrm{PS}}$ and $(r-i)_{\mathrm{PS}}$
with the photometric errors and uncertainties of the CTGs and CTIs.

The same as color derivations,
we used the linear relationship to derive the magnitudes as follows:
\begin{eqnarray}
    g_m^n &= g_{\mathrm{inst}, m}^n + \mathrm{CT}_{g} (g - r)_{\mathrm{PS}, m}^n  + \mathrm{Z}_g^n, \label{eq:g_CT} \\
    r_m^n &= r_{\mathrm{inst}, m}^n + \mathrm{CT}_{r} (g - r)_{\mathrm{PS}, m}^n  + \mathrm{Z}_r^n, \label{eq:r_CT} \\
    i_m^n &= i_{\mathrm{inst}, m}^n + \mathrm{CT}_{i} (g - r)_{\mathrm{PS}, m}^n  + \mathrm{Z}_i^n, \label{eq:i_CT}
\end{eqnarray}
where
$g_m^n$, $r_m^n$, and $i_m^n$ are magnitudes in the Pan-STARRS system of 
$m$-th object on the $n$-th frame, 
$g_{\mathrm{inst}, m}^n$, $r_{\mathrm{inst}, m}^n$, and $i_{\mathrm{inst}, m}^n$
are instrumental magnitudes of $m$-th object on the $n$-th frame,
$\mathrm{CT}_{g}$, $\mathrm{CT}_{r}$, and $\mathrm{CT}_{i}$,
are color terms (CTs) of $g$, $r$, and $i$-band magnitudes, respectively,
and $\mathrm{Z}_g^n$, $\mathrm{Z}_r^n$, and $\mathrm{Z}_i^n$ 
are zero points of $g$, $r$, and $i$-band magnitudes on the $n$-th frame, respectively.
The same as colors,
we firstly derived CTs at the night 
while shifting the photometric results of the reference stars during the night frame by frame
so that the atmospheric variations in instrumental magnitudes and Zs were canceled out.
Then, the Zs in each frame were calculated with the derived CTs with equations (\ref{eq:g_CT})--(\ref{eq:i_CT}).
We computed the propagated uncertainties of 
$g$, $r$, and $i$-band magnitudes
with the photometric errors and uncertainties of $(g-r)_{\mathrm{PS}}$, CTs and Zs.

\subsubsection{Periodic analysis}\label{subsub:peri}
We used the Lomb-Scargle method for periodic analysis
\citep{Lomb1976, Scargle1982, VanderPlas2018}.
The detail of the periodic analysis is summarized in \citet{Beniyama2022}.
The purpose of the periodic analysis in this study is not to
derive the rotation period of Phaethon, but to create fitting curves
to estimate the mean magnitude, which is essential for the phase curve fitting.
Thus, we referred to the reported rotation period of 
Phaethon $P=3.603957\pm0.000001$ h \citep{Kim2018, Hanus2018}
and fixed it to determine the Fourier coefficients of the model curve.
Since the Phaethon's sidereal and synodic periods during our observations 
have little difference ($\sim1.2\,\mathrm{s}$, \cite{Harris1984}),
we referred to the Phaethon's sidereal period derived in the previous studies.

\subsubsection{Phase curve fitting}\label{subsub:pc}
The reduced $g$, $r$ and $i$-band magnitudes,
$g_\mathrm{red}$, $r_\mathrm{red}$, and $i_\mathrm{red}$,
were obtained 
by scaling the distances from both Earth and Sun to Phaethon to 1\,au as follows:
\begin{eqnarray}
    g_{\mathrm{red}}=g-5\log_{10}{\Delta r_\mathrm{h}},\\
    r_{\mathrm{red}}=r-5\log_{10}{\Delta r_\mathrm{h}},\\
    i_{\mathrm{red}}=i-5\log_{10}{\Delta r_\mathrm{h}},
\end{eqnarray}
where 
$\Delta$ and $r_\mathrm{h}$ are geocentric and heliocentric distances in the unit of au
at the time of observation.
With the same Fourier coefficients of the model curve created in section \ref{subsub:peri} 
except for the constant term (i.e., mean magnitude),
we searched the optimal constant term to minimize the residuals between the model and the observations.
The averages of the reduced magnitudes 
$g_\mathrm{red, mean}$, $r_\mathrm{red, mean}$, and $i_\mathrm{red, mean}$, 
were derived as constant terms of the Fourier series 
of the shifted model curve (see dotted lines in figure \ref{fig:alllcs_abs} in the Appendix \ref{app:pc}).

The $g_\mathrm{red, mean}$ and $r_\mathrm{red, mean}$ were
converted to $R$-band magnitudes using the equation in \citet{Tonry2012}
to refer to \citet{Ansdell2014} for data at other phase angles, 
in which Phaethon was observed with $R$-band filter (see section \ref{sub:res_pc} for detail).
In section \ref{sub:res_pc},
we used two types of models in phase curve fitting 
to derive the absolute magnitude and the slope parameter.

The posterior distributions of the parameters were estimated using the Monte Carlo technique.
We created 1000 phase curves by randomly sampling the observed data points 
assuming that every point follows a normal distribution.
We adopted the standard deviations 
as the uncertainties of the parameters.
The median (50th percentile) and 95 \% highest density interval (HDI) values
were obtained.

\section{Results}\label{sec3}

\subsection{The comet-like activity}
\begin{figure}
\includegraphics[width=80mm]{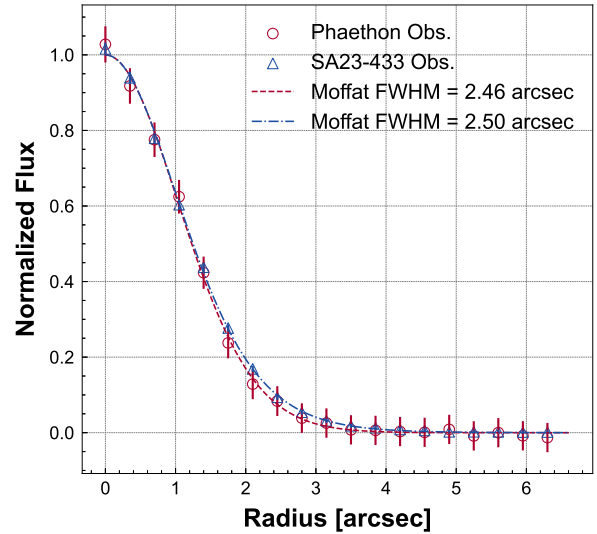}
\caption{%
    Radial profiles of Phaethon (circles) and SA23-433 (triangles) 
    in the $r$-band images in October 28th, 2021.
    The profiles were normalized to the peaks of the Moffat fits (dashed and dot-dashed lines).
    Bars indicate the 1-$\sigma$ uncertainties.}
\label{fig:psf}
\end{figure}

The comet-like activity near the perihelion,
possibly related to the dust production,
was reported for Phaethon 
(Jewitt \& Li\,\yearcite{Jewitt2010}; \cite{Jewitt2013a}; Li \& Jewitt\,\yearcite{Li2013}; Hui \& Li\,\yearcite{Hui2017}).
Though Phaethon was far from its perihelion at the time of our observing runs,
we searched the comet-like activity of Phaethon in the 
$r$-band image in October 28th, 2021.
A comparison of the radial profile of Phaethon with that of a standard star SA23-433 \citep{Landolt2013} is shown in figure \ref{fig:psf}.
The standard star SA23-433 was observed in the sidereal tracking mode
before the Phaethon's image.
The exposure starting times of the images of Phaethon and SA23-433 are
2021-10-28 16:06:55 and 2021-10-28 16:11:38 (UT), respectively.
Both images were obtained in similar low-airmass conditions ($\sim1.02$).
The fluxes were averaged in the azimuthal direction from the centroids of Phaethon and SA23-433.
The uncertainties of the fluxes consist of the background noise and the Poisson noise in each annulus.
Moffat functions were used to fit both radial profiles to derive the FWHMs of the PSFs.
The derived FWHMs were 2.46 arcsec and 2.50 arcsec for Phaethon and SA23-433, respectively.
The similarity of the radial profile of Phaethon and SA23-433
implies that no noticeable comet-like activity was found in the 2021 apparition.

\subsection{Lightcurves}\label{sub:lc}
\begin{figure}
\includegraphics[width=80mm]{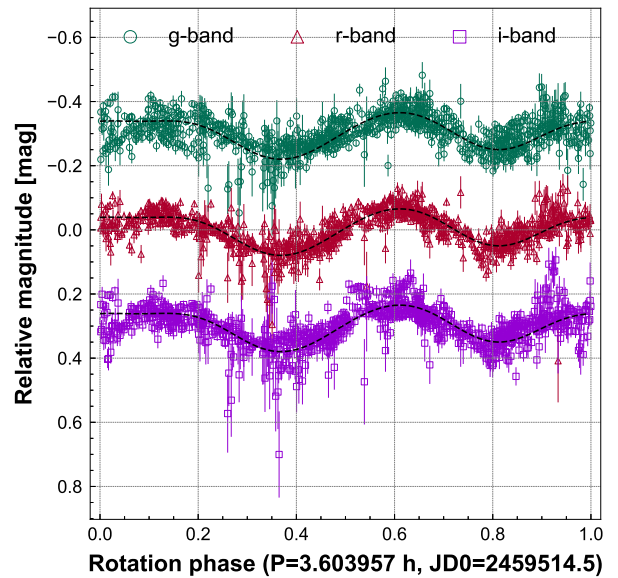}
\caption{%
    Phased relative lightcurves of Phaethon in nine nights.
    From top to bottom, g (circles), r (triangles), and i-band (squares) 
    relative lightcurves in each day are presented.
    Lightcurves are folded by the reported period of $3.603957\,\mathrm{h}$ and 
    phase zero is set to 2459514.5\,JD.
    Bars indicate the 1-$\sigma$ uncertainties.
    Fitting model curves are shown by dashed lines.
    Phased lightcurves of Phaethon in each day are attached in figure \ref{fig:alllcs_abs}, see the Appendix \ref{app:pc}.}
\label{fig:mag_lc}
\end{figure}

The folded lightcurves with $P=3.603957\,\mathrm{h}$ \citep{Kim2018, Hanus2018} in nine nights are shown in figure \ref{fig:mag_lc}.

The fitting model curves (i.e., Fourier coefficients) were obtained 
using the $g$-band lightcurves on November 24, 25, and 26, adopting the number of harmonics $n=4$.
Each lightcurve was shifted to minimize the residuals between the model and the observations
considering the constant term (i.e., mean magnitude) as a free parameter.
We defined a peak-to-trough variation as a lightcurve amplitude $\Delta m$.
Assuming the asteroid is 
a triaxial ellipsoid with axial lengths of $a$, $b$, and $c$ ($a > b > c$)
and rotating along the $c$-axis with an aspect angle of $90^{\circ}$,
a lower limit of axial ratio $a/b$ is estimated as follows:
\begin{eqnarray}
    \frac{a}{b} \geq
    10^{0.4\Delta m(\alpha)/(1+s\alpha)},
    \label{eq:ab}
\end{eqnarray}
where $\Delta m(\alpha)$ is a lightcurve amplitude at a phase angle of $\alpha$
and $s$ is a slope depending on the taxonomic type of the asteroid \citep{Bowell1989}.
We assumed $s=0.015$, a typical value for C-type asteroids \citep{Zappala1990}.
The lightcurve amplitude of the fitting model curve in 
the $g$-band was derived as 0.15 mag,
which corresponds to $a/b\geq1.13$ at $\alpha= 9.1^{\circ}$.

\subsection{Surface colors} \label{subsec:color}
\begin{figure*}
\includegraphics[width=160mm]{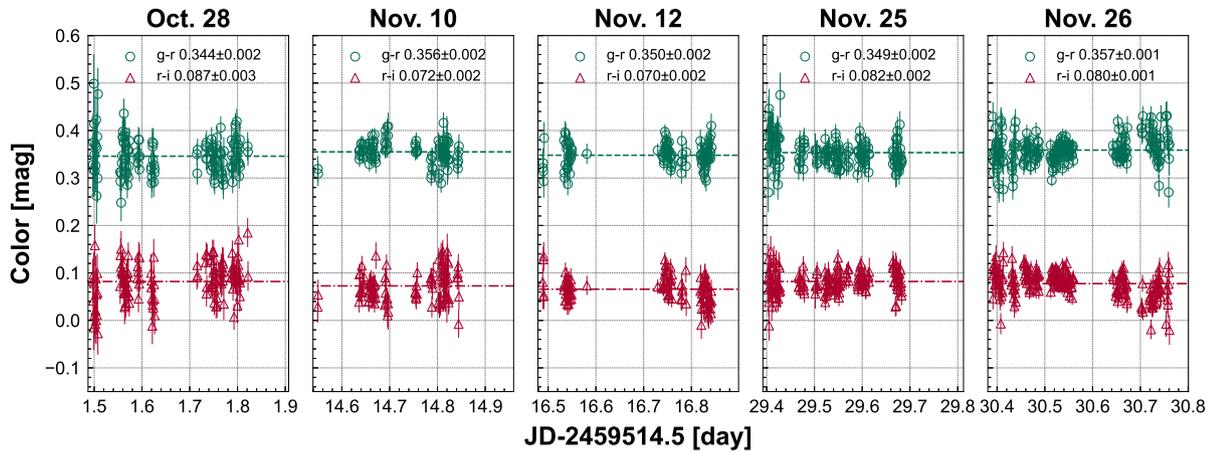}
\caption{%
    Colors of Phaethon in five photometric nights.
    $g-r$ and $r-i$ colors are shown by circles and triangles, respectively.
    Weighted averages of $g-r$ and $r-i$ colors in each night are presented with
    dashed and dash-dotted horizontal lines, respectively. 
    Bars indicate the 1-$\sigma$ uncertainties.}
\label{fig:color_lc}
\end{figure*}

The colors in five photometric nights, 
October 28, November 10, 12, 25, and 26, are shown in figure \ref{fig:color_lc}.
We calculated the weighted averages of the colors in each night.
Here we presume that the systematic uncertainty of colors derived by our observations 
is 0.02 mag based on the photometry of reference stars (see the Appendix \ref{app:photcon}).
The systematic uncertainties dominate over the statistical uncertainties in derived colors.
Then, the weighted average of colors were $g-r=0.35\pm0.02$ and $r-i=0.08\pm0.02$.
The color corresponds to $V-R=0.36\pm0.02$ in the Johnson system \citep{Tonry2012}.

We summarized the derived surface colors with those reported in previous studies
in table \ref{tab:color}.
To compare the colors obtained in different filter systems, 
the color transformations between the Pan-STARRS and the Johnson systems were 
performed with the equations in \citet{Tonry2012}.
Since \citet{Dundon2005} in 2004, \citet{Jewitt2013b}, 
and \citet{Ansdell2014} did not observe Phaethon in the $I$-band, 
we derived only the $g-r$ color in the Pan-STARRS system for them.

\begin{table*}
      \tbl{Comparison of Phaethon's surface colors.\footnotemark[$*$]}{%
         % 8 columns
         \begin{tabular}{lcccccc}
           \hline
       References            & Obs. Date             & $g-r$                   & $r-i$                    & $B-V$                     & $V-R$                    &  $R-I$                    \\ 
                             & (UTC)                 & (mag)                   & (mag)                    & (mag)                     & (mag)                    &  (mag)                    \\ 
       \hline
     \citet{Dundon2005} & 1996 Nov 12& $ 0.37 \pm 0.03 $& $ - $& $ \mathbf{0.618 \pm 0.005}$& $ \mathbf{0.347 \pm 0.004}$& $ - $\\
\citet{Dundon2005} & 1997 Nov 22& $ 0.40 \pm 0.03 $& $ 0.08 \pm 0.03 $& $ \mathbf{0.650 \pm 0.004}$& $ \mathbf{0.295 \pm 0.002}$& $ \mathbf{0.320 \pm 0.003}$\\
\citet{Dundon2005} & 2004 Nov 19& $ 0.34 \pm 0.03 $& $ - $& $ \mathbf{0.587 \pm 0.005}$& $ \mathbf{0.349 \pm 0.003}$& $ - $\\
Kasuga \& Jewitt\,(\yearcite{Kasuga2008}) & 2007 Sep 04& $ 0.36 \pm 0.03 $& $ 0.03 \pm 0.05 $& $ \mathbf{0.61 \pm 0.01}$& $ \mathbf{0.34 \pm 0.03}$& $ \mathbf{0.27 \pm 0.04}$\\
\citet{Jewitt2013b} & 2010 Sep 10& $ 0.42 \pm 0.03 $& $ - $& $ \mathbf{0.67 \pm 0.02}$& $ \mathbf{0.32 \pm 0.02}$& $ - $\\
\citet{Ansdell2014} & 1997 Nov 12& $ 0.34 \pm 0.03 $& $ - $& $ \mathbf{0.58 \pm 0.01}$& $ \mathbf{0.34 \pm 0.02}$& $ - $\\
\citet{Ansdell2014} & 1997 Nov 22& $ 0.33 \pm 0.03 $& $ - $& $ \mathbf{0.57 \pm 0.01}$& $ \mathbf{0.36 \pm 0.01}$& $ - $\\
\citet{Ansdell2014} & 1995 Jan 04& $ 0.28 \pm 0.03 $& $ - $& $ \mathbf{0.52 \pm 0.01}$& $ \mathbf{0.33 \pm 0.01}$& $ - $\\
\citet{Lee2019} & 2017 Nov 11--13& $ 0.39 \pm 0.03 $& $ 0.07 \pm 0.04 $& $ \mathbf{0.64 \pm 0.02}$& $ \mathbf{0.34 \pm 0.02}$& $ \mathbf{0.31 \pm 0.03}$\\
\citet{Tabeshian2019} & 2017 Dec 11--19& $ 0.45 \pm 0.03 $& $ 0.03 \pm 0.03 $& $ \mathbf{0.702 \pm 0.004}$& $ \mathbf{0.309 \pm 0.003}$& $ \mathbf{0.266 \pm 0.004}$\\
\citet{Lin2020} & 2017 Oct/Nov/Dec& $ 0.38 \pm 0.04 $& $ 0.10 \pm 0.03 $& $ \mathbf{0.633 \pm 0.036}$& $ \mathbf{0.336 \pm 0.011}$& $ \mathbf{0.334 \pm 0.015}$\\
Sergeyev \& Carry\,(\yearcite{Sergeyev2021a})$^{\dag}$ & 2009 Jan 26& $ 0.32 \pm 0.03 $& $ 0.07 \pm 0.03 $& $ 0.56 \pm 0.05$& $ 0.34 \pm 0.03$& $ 0.30 \pm 0.04$\\
This work & 2021 Oct 28& $ \mathbf{0.34 \pm 0.02}$& $ \mathbf{0.09 \pm 0.02}$& $ 0.59 \pm 0.04 $& $ 0.35 \pm 0.02 $& $ 0.32 \pm 0.03 $\\
This work & 2021 Nov 10& $ \mathbf{0.36 \pm 0.02}$& $ \mathbf{0.07 \pm 0.02}$& $ 0.60 \pm 0.04 $& $ 0.36 \pm 0.02 $& $ 0.31 \pm 0.03 $\\
This work & 2021 Nov 12& $ \mathbf{0.35 \pm 0.02}$& $ \mathbf{0.07 \pm 0.02}$& $ 0.60 \pm 0.04 $& $ 0.36 \pm 0.02 $& $ 0.31 \pm 0.03 $\\
This work & 2021 Nov 25& $ \mathbf{0.35 \pm 0.02}$& $ \mathbf{0.08 \pm 0.02}$& $ 0.60 \pm 0.04 $& $ 0.36 \pm 0.02 $& $ 0.32 \pm 0.03 $\\
This work & 2021 Nov 26& $ \mathbf{0.36 \pm 0.02}$& $ \mathbf{0.08 \pm 0.02}$& $ 0.60 \pm 0.04 $& $ 0.36 \pm 0.02 $& $ 0.32 \pm 0.03 $\\
\hline
          \end{tabular}
    }\label{tab:color}
    \begin{tabnote}
        \footnotemark[$*$] $g-r$, $r-i$ (Pan-STARRS), $B-V$, $V-R$, and $R-I$ (Johnson) colors are listed with their references. 
        Colors derived without any transformation are shown in bold character.
        Colors derived after transformations based on \citet{Tonry2012} are presented in normal character.\\
        \footnotemark[$\dag$] The Pan-STARRS and Johnson colors are converted from the SDSS colors.
    \end{tabnote}\end{table*}

\subsection{Phase curve}\label{sub:res_pc}
\begin{figure}
\includegraphics[width=80mm]{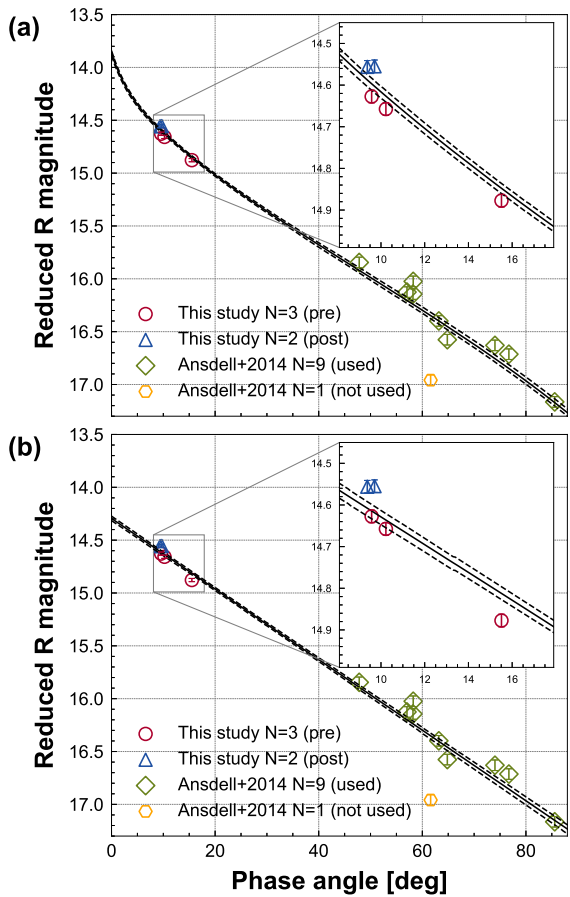}
\caption{%
    Phase curves of Phaethon with
    (a) the $H$-$G$ and (b) the linear models.
    Mean reduced $R$-band magnitudes in 2021 pre-opposition,
    2021 post-opposition, and \citet{Ansdell2014} 
    are presented as circles, triangles, and diamonds, respectively.
    Removed data in \citet{Ansdell2014} are presented as Hexagons 
    (see text for detail).
    Bars indicate the 1-$\sigma$ uncertainties.
    Median (50th percentile) of fitting model curves 
    are presented by solid lines.
    Uncertainty envelopes representing 
    the 95 \% HDI values
    are shown by dashed lines.}
\label{fig:phasecurve}
\end{figure}

We successfully observed Phaethon at lower phase angles down to about $9^{\circ}$.
The phase curves of Phaethon are shown in figure \ref{fig:phasecurve}.
As in subsection \ref{subsec:color}, we limited the results obtained in the five photometric nights.

We combined our observational results with 
those reported in \citet{Ansdell2014}, not \citet{Tabeshian2019}.
This is because we regard the average of the $B-V$ color in \citet{Tabeshian2019} are different from 
the colors reported in other studies (table \ref{tab:color}).
The data set in \citet{Ansdell2014} 
obtained in November, 2004 and December, 2013 
are not used in this paper
since those lightcurves are relative rather than absolute without absolute 
calibrations due to the lack of field stars.
We note that observational results in 2013 Nov 13 in \citet{Ansdell2014}
are also removed by eye and not used for phase curve fittings 
since the average of the reduced $R$-band magnitude is far from other observations.

The uncertainties of $g_\mathrm{red, mean}$ and $r_\mathrm{red, mean}$
are given as the standard error of the weighted average of the lightcurves.
These uncertainties are much smaller than 
those of the mean $R$-band magnitudes of our observations ($\sim$0.015)
due to the uncertainties in the magnitude transformation between the Pan-STARRS and the Johnson systems.
This is the same for the observation results in \citet{Ansdell2014},
but the propagated uncertainties of the mean $R$-band magnitudes are smaller than 
ours since there is no magnitude transformation.
Taking into account the fact that the lightcurve amplitudes of Phaethon are about 0.1--0.2 mag 
and not all lightcurves in \citet{Ansdell2014} fully covered the rotation phase, 
we set the uncertainties of the mean reduced $R$-band magnitudes in \citet{Ansdell2014} to 0.05 mag.
We obtained the mean $R$-band magnitudes as 
$R=14.88\pm0.02\,\,\mathrm{mag}$ at $\alpha=15.5^{\circ}$, 
$R=14.66\pm0.02\,\,\mathrm{mag}$ at $\alpha=10.2^{\circ}$, 
$R=14.63\pm0.02\,\,\mathrm{mag}$ at $\alpha=9.6^{\circ}$, 
$R=14.56\pm0.02\,\,\mathrm{mag}$ at $\alpha=9.4^{\circ}$,
and 
$R=14.55\pm0.02\,\,\mathrm{mag}$ at $\alpha=9.7^{\circ}$,
where $\alpha$ were the mean solar phase angles at the nights.

First, we derived an absolute magnitude in $R$-band $H_\mathrm{R}$ and a slope parameter $G$ 
with the $H$-$G$ model \citep{Bowell1989}:
\begin{eqnarray}
    H_\mathrm{R}(\alpha) = H_\mathrm{R} - 2.5 \log_{10}{((1-G)\Phi_1(\alpha)+G\Phi_2(\alpha)}),
\end{eqnarray}
where $\alpha$ is a phase angle. 
$\Phi_1$ and $\Phi_2$ are phase functions written as follows with a basic function $W$:
\begin{eqnarray}
    \Phi_1(\alpha) &= W\left(1-\frac{0.986\sin{\alpha}}{0.119+1.341\sin{\alpha}-0.754\sin^2{\alpha}}\right) \nonumber \\
                   &+ (1-W) \exp{\left(-3.332\tan^{0.631}{\frac{\alpha}{2}}\right)},\\
    \Phi_2(\alpha) &= W\left(1-\frac{0.238\sin{\alpha}}{0.119+1.341\sin{\alpha}-0.754\sin^2{\alpha}}\right) \nonumber \\
                   &+ (1-W) \exp{\left(-1.862\tan^{1.218}{\frac{\alpha}{2}}\right)},\\
    W              &= \exp{\left(-90.56\tan^2{\frac{\alpha}{2}}\right)}.
\end{eqnarray}
We derived 
$H_\mathrm{R}=13.87\pm0.01$ and $G=0.040\pm0.008$ from model
fitting with $H$-$G$ model in panel (a) of figure \ref{fig:phasecurve}.
$H_\mathrm{R}$ were converted to 
$H_\mathrm{V}=14.23\pm0.02$
using the $V-R$ color derived in subsection \ref{subsec:color}.
The phase reddening effect is ignored in this study since 
the colors of B-type asteroids in Bus or Bus-DeMeo taxonomy \citep{DeMeo2009} 
are known to be almost constant at the phase angle $\alpha \leq 70^{\circ}$
\citep{Lantz2018}.

Second, we used the linear model:
\begin{eqnarray}
    H_\mathrm{R}(\alpha) = H_\mathrm{R, lin} + a_\mathrm{lin}\alpha,
\end{eqnarray}
where, 
$H_\mathrm{R, lin}$ is an absolute magnitude in the linear model
and $a_\mathrm{lin}$ is a slope of the fitting curve.
In panel (b) of figure \ref{fig:phasecurve},
we derived 
$a_\mathrm{lin}=0.0334\pm0.0003$ and $H_\mathrm{R, lin}=14.29\pm0.01$ with the linear model,
which corresponds to 
$H_\mathrm{V}=14.65\pm0.02$.

We also tried the phase curve fitting with Shevchenko's model \citep{Shevchenko1996},
but the fitting curve showed a strong opposition surge.
This is possibly due to the lack of observations at lower phase angles ($\alpha \leq 9^{\circ}$). 
Thus, we did not derive the absolute magnitude with the Shevchenko model.

\section{Discussion}\label{sec4}

\subsection{Updated absolute magnitude}
The opposition surge is seen in phase curves with the $H$-$G$ model in panel (a) of figure \ref{fig:phasecurve}.
This is not the case for the linear model by definition in panel (b) of figure \ref{fig:phasecurve}.
Whether the opposition surge exists in Phaethon's phase curve
is unclear even with our latest results ($\alpha \geq 9^{\circ}$).
\citet{Devogele2020} found that the phase curve of 2005 UD,
which may have the same parent body as Phaethon 
(\cite{Ohtsuka2006}; Jewitt \& Hsieh\,\yearcite{Jewitt2006}), 
shows very little or no opposition surge.
Although, Phaethon and 2005 UD may have similar properties by chance \citep{Ryabova2019, Kareta2021}.
In that case, the opposition surge may exist in Phaethon's phase curve.
We conservatively derived a possible absolute magnitude range as 
$H_\mathrm{V}=$14.18--14.66 with $H_R$, $H_\mathrm{R, lin}$, and their uncertainties
considering both cases with and without the opposition surge.
The comparison with the phase curve by in-situ observations of DESTINY$^{+}$ is awaited
to investigate the existence of the opposition surge.

\subsection{Diameter}
We combined our $H_\mathrm{V}$ with the latest geometric albedo 
$p_\mathrm{V}=$0.08--0.13 derived from polarimetric observations \citep{Geem2022}
to estimate the rotation averaged area-equivalent diameter of Phaethon.
The reasons to refer to \citet{Geem2022} are
that the albedos can be derived directly from the polarimetric parameters 
using an empirical equation and they carefully considered all
possible uncertainties of constants in the empirical equation. 
The diameter of an asteroid is estimated with $H_\mathrm{V}$ and $p_\mathrm{V}$
using the following equation (Fowler\,\&\,Chillemi \yearcite{Fowler1992}):
\begin{eqnarray}
    D=\frac{1329}{\sqrt{p_\mathrm{V}}}\times10^{-H/5}\,\,\mathrm{km}.
    \label{eq:D}
\end{eqnarray}
The Phaethon's diameter
was estimated to be within the range of 
$5.22 \leq D \leq 6.74\,\,\mathrm{km}$ with the $H$-$G$ model
and $4.30 \leq D \leq 5.56\,\,\mathrm{km}$ with the linear model.

In table \ref{tab:D}, we compared the derived Phaethon's diameter 
with those estimated by other techniques:
radar observations, TPMs, and occultation observations.
\citet{Taylor2019} reported the volume-equivalent diameters as 6.2\,km and 5.5\,km
using the Doppler-only echo spectra with the spherical and top-shaped assumptions, 
respectively.
Also, the mean equatorial diameter was derived as 6.25\,km using
the Range-Doppler technique.
\authorcite{Hanus2016} (2016, 2018) derived a convex shape model of Phaethon
and estimated the volume-equivalent diameter as $5.1\pm0.2$ km by the TPM with thermal infrared data taken by 
IRAS, UKIRT \citep{Green1985}, 
and Spitzer Space Telescope in a combination with optical lightcurves.
They assumed $H_\mathrm{V} = 14.31$ and $G = 0.15$ in their TPM.
% AKARI is not public.
\citet{Masiero2019} estimated the volume-equivalent diameter as $4.6^{+0.2}_{-0.3}$ km 
with the spherical assumption
by the TPM with thermal infrared data taken by WISE.
They assumed $H_\mathrm{V} = 14.31\pm0.03$ in their TPM.
Recently, 
\citet{MacLennan2022b} derived a volume-equivalent diameter of $5.4\pm0.1$\,km 
using a TPM which incorporates AKARI observations \citep{Usui2011} 
in addition to the data used in all the prior studies with an accurate non-convex shape model.
As for occultation observations, \citet{Devogele2020} combined Phaethon's 
convex shape model in \citet{Hanus2018} and 
occultation chords in the USA on July 29, 2019.
The volume-equivalent diameter is derived as $5.2\pm0.1$\,km.
\citet{Yoshida2022} reported that the projected size of Phaethon is 
$6.12\pm0.07\,\,\mathrm{km}\times 4.14\pm0.07\,\,\mathrm{km}$
by stellar occultation observations in Japan on October 13, 2021.
An area-equivalent diameter of a projected ellipse is estimated to be $5.03\pm0.07$\,km.

\begin{table*}
      \tbl{Comparison of Phaethon's Diameters.%\footnotemark[$*$]
      }{%
         \begin{tabular}{lllc}
           \hline
      References                                        & Methods         & Shape  & $D$              \\
                                                        &                 &        & (km)              \\ \hline
      \citet{Taylor2019} & Doppler-only & sphere & 6.2\footnotemark[$*$] \\
\citet{Taylor2019} &  Doppler-only & top-shape & 5.5\footnotemark[$*$] \\
\citet{Taylor2019} &   Range-Doppler& - &6.25\footnotemark[$\dag$] \\
\authorcite{Hanus2016} (2016, 2018) &   TPM & convex shape & $5.1\pm{0.2}$\footnotemark[$*$] \\
\citet{Masiero2019} &   TPM & sphere & $4.6^{+0.2}_{-0.3}$\footnotemark[$*$] \\
\citet{MacLennan2022b} &   TPM & non-convex shape & $5.4\pm0.1$\footnotemark[$*$] \\
\citet{Devogele2020} & Occultation & convex shape & $5.2\pm0.1$\footnotemark[$*$] \\
\citet{Yoshida2022} & Occultation &ellipse& $5.03\pm0.07$\footnotemark[$\ddag$] \\
 This study $H$-$G$ model + Geem et al. (2022) & Photometry \& Polarimetry & - &5.22--6.74 \footnotemark[$\S$] \\
 This study linear model + Geem et al. (2022) & Photometry \& Polarimetry & - &4.30--5.56 \footnotemark[$\S$] \\\hline
            \end{tabular}
      }\label{tab:D}
      \begin{tabnote}
          \footnotemark[$*$] Volume-equivalent diameter\\
          \footnotemark[$\dag$] Mean equatorial diameter\\
          \footnotemark[$\ddag$] Area-equivalent diameter of a projected ellipse\\
          \footnotemark[\S] Rotation averaged area-equivalent diameter
      \end{tabnote}\end{table*}

The diameter derived with the $H$-$G$ model in this study 
falls into the larger range of diameters.
This is consistent with radar and occultation observations.
On the other hand, the diameter derived with the linear model
falls into the smaller range of diameters, which is consistent with those derived by the TPMs.

\subsection{Surface inhomogeneity}
We conducted simultaneous tricolor observations of Phaethon for the first time.
The simultaneity in the multicolor photometry is crucial 
to investigate the surface inhomogeneity of the asteroid
since it can reduce the effects of atmospheric variations on the measurements.

We referred to two published observational studies on surface spectroscopic variations of Phaethon.
By multicolor photometry,
\citet{Tabeshian2019} reported a $B–V$ color variation of about 0.1 mag 
with observations from December 11 to 19 in 2017.
They interpreted that the color variation with sub-observer latitude 
may be associated with craters on the surface since 
they observed both southern and northern parts in the 2017 apparition.
\citet{Lin2020} detected a $R-I$ color variation of 0.037\,mag in 
rotation phase with observations from October 6 to December 21 in 2017.
They interpreted that the color variation may be caused by the inhomogeneous surface.

\begin{figure*}
\includegraphics[width=160mm]{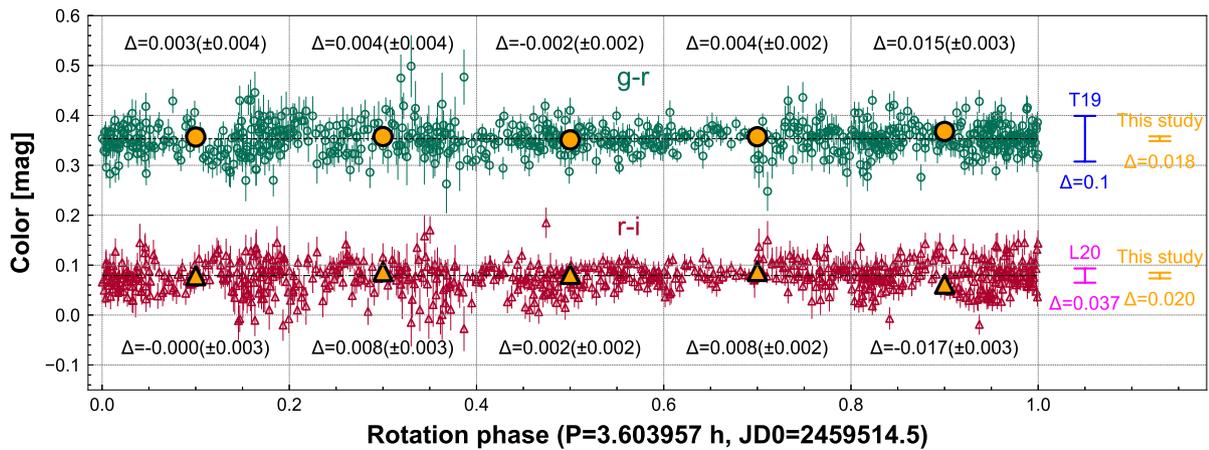}
\caption{%
    Phased colors of Phaethon in five photometric nights.
    $g-r$ and $r-i$ colors are shown by open circles and open triangles, respectively.
    Weighted average $g-r$ and $r-i$ colors in each night are presented with
    dashed and dash-dotted horizontal lines, respectively. 
    Weighted average $g-r$ and $r-i$ colors per 0.2 phase are shown by filled circles and filled triangles
    with derivations from the global average colors.
    Values in the parentheses are uncertainties of the binned colors.
    Range corresponding to the color variations reported in \authorcite{Tabeshian2019} (2019, T19), \authorcite{Lin2020} (2020, L20), and this study are presented by arrows on the right.}
\label{fig:phasedcolor}
\end{figure*}

We divided the phased colors into five sections equally
and calculated the weighted mean colors as shown in figure \ref{fig:phasedcolor}.
The largest deviation from the global mean values 
are 0.018 mag and 0.020 mag in the $g-r$ and $r-i$ colors, respectively,
considering the uncertainties.
Color variations in the 2021 apparition are smaller than those reported in previous studies.

Phaethon's latitudinal surface inhomogeneity can probably explain this inconsistency.
As is often the case in near-Earth asteroids,
the aspect angles are different in each apparition and sometimes even in the same apparition.
We mainly observed the slightly northern part of Phaethon in our observing runs;
sub-observer latitudes were calculated with the pole orientation 
$\lambda=316^{\circ}$ and $\beta=-48.7^{\circ}$ \citep{MacLennan2022b}
as 29.7 to 29.2$^{\circ}$ between Oct. 27 and 29, 
25.6 to 24.8$^{\circ}$ between Nov. 10 and 12, 
and 19.5 to 18.5$^{\circ}$ between Nov. 24 and 26. 
On the other hand, the aspect angle was changed a lot in the 2017 apparition.
Thus, our observational result suggests that Phaethon's northern hemisphere is almost homogeneous 
compared with the southern/northern latitudinal difference seen in \citet{Tabeshian2019}.
This may be relevant to the latitudinal heterogeneity of the surface grain sizes on Phaethon \citep{MacLennan2022b}.

Both the latitudinal surface inhomogeneity and our homogeneous color are 
not necessarily inconsistent with the color variation in the rotation phase reported in \citet{Lin2020}.
If the colors 
in the former (0.0--0.5) and the latter (0.5--1.0) rotation phases in \citet{Lin2020} 
were derived using the data obtained in the different periods (i.e., aspect angles),
this is consistent with the latitudinal surface inhomogeneity.
Since there is no detailed information for observations in \citet{Lin2020},
such as the observation date and time of each data,
further discussion is beyond the scope of this paper.
We note that there is a possibility that the reported surface color variation is affected by observational artifacts.
The situation is the same for 2005 UD.
\citet{Kinoshita2007} detected the surface color variation on 2005 UD by multicolor photometry in 2005.
However, \citet{Devogele2020} did not detect any color variations on 2005 UD by spectroscopy on two nights in 2018.
For reliable detections of surface color variation, 
independent observations (e.g., optical and NIR, \cite{Lopez-Oquendo2022})
and/or observations in multiple rotation phases
are desired in addition to careful data reduction.
As for Phaethon,
the DESTINY$^{+}$ mission might put an end to the discussion related to the 
surface inhomogeneity on Phaethon.

\section{Conclusions}\label{sec5}
We conducted simultaneous tricolor photometry for the first time for 
the DESTINY$^{+}$ mission target asteroid (3200) Phaethon.
By observations at the lower phase angles down to $\sim$9$^{\circ}$,
we updated an absolute magnitude $H_\mathrm{V}$ and a slope parameter $G$ 
as $H_\mathrm{V}=14.23\pm0.02$ and $G=0.040\pm0.008$ with the $H$-$G$ model.
Assuming geometric albedo derived by polarimetry,
Phaethon's diameter was estimated to be in the range of 
5.22--6.74\,km,
which is consistent with diameters derived from other techniques 
such as radar and occultation observations.
We derived $H_\mathrm{V}=14.65\pm0.02$ with the linear model,
which corresponds to a diameter range of 4.30--5.56\,km.
The diameters by thermophysical modelings should be updated using our derived $H_\mathrm{V}$ and $G$.
No noticeable rotational spectral variation was found 
in our simultaneous multicolor lightcurves in 2021.
Our observational result suggests Phaethon's northern hemisphere is more homogeneous than
the southern/northern latitudinal difference reported in previous studies.

%%% Acknowledgments
\begin{ack}
The authors are grateful to our reviewer Eric MacLennan 
for helpful comments on the first version of this manuscript.
J. B. would like to express my gratitude to the Iwadare Scholarship Foundation 
and the Public Trust Iwai Hisao Memorial Tokyo Scholarship Fund for the grants.
This work is supported in part by the Optical and Near-Infrared Astronomy Inter-University Cooperation Program (OISTER), 
the Ministry of Education, Culture, Sports, Science and Technology (MEXT) of Japan,
JST SPRING, Grant Number JPMJSP2108, and the UTEC UTokyo Scholarship.
The authors thank the TriCCS developer team 
(which has been supported by the JSPS KAKENHI grant Nos. JP18H05223, 
JP20H00174, and JP20H04736, and by NAOJ Joint Development Research).
\end{ack}

%%% Appendix
\appendix
\section{Phased lightcurves} \label{app:pc}
\begin{figure*}
\includegraphics[width=16cm]{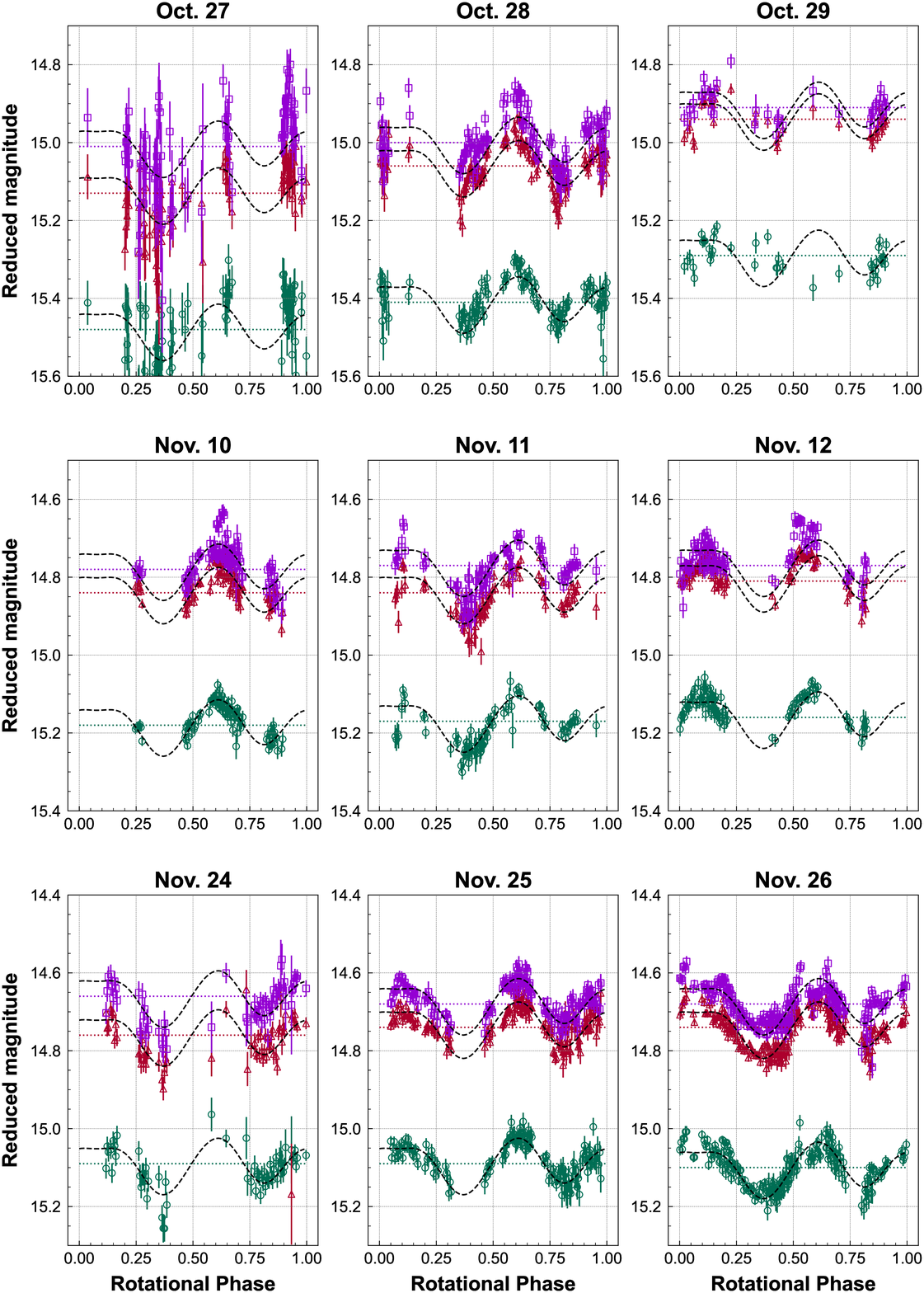}
\caption{%
    Phased lightcurves of Phaethon in nine days.
    From top to bottom, g (circles), r (triangles), and i-band (squares) 
    lightcurves on different days are presented. Bars indicate the 1-$\sigma$ uncertainties.
    Model curves adopting the number of harmonics $n=4$ are shown by dashed lines.
    Mean reduced magnitudes are given as constant terms of the model curves as 
    presented by dotted lines.}
\label{fig:alllcs_abs}
\end{figure*}

Phased lightcurves in each day are presented here in figure \ref{fig:alllcs_abs}.

\section{Photometric conditions} \label{app:photcon}
\begin{figure*}
\includegraphics[width=14cm]{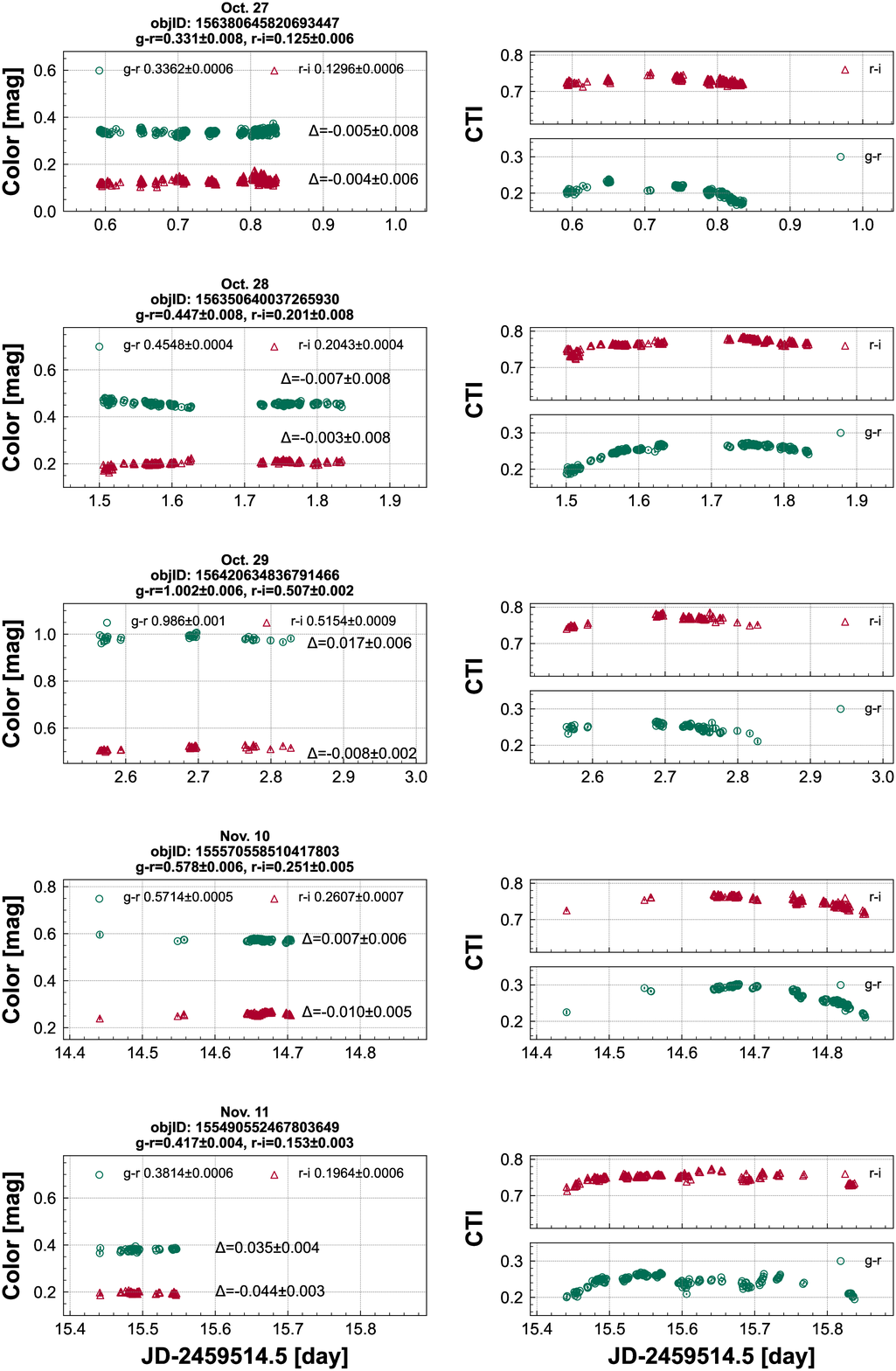}
\caption{%
    Colors of reference stars on October 26, 27, 28, November 10, and 11 (left).
    Object name (\texttt{objID}) and colors in the Pan-STARRS DR2 are written on the top of the panel.
    $g-r$ and $r-i$ colors are shown by circles and triangles, respectively.
    Bars indicate the 1-$\sigma$ uncertainties.
    The differences $\Delta$, colors in Pan-STARRS DR2 - those derived here,
    are given.
    The number of observations of reference stars differs from star to star 
    due to the change of the field of view.
    Time-series CTIs on October 26, 27, 28, November 10, and 11 (right).
    CTI$_{g-r}$ and CTI$_{r-i}$ are shown by circles and triangles, respectively.
    Bars indicate the 1-$\sigma$ uncertainties.}
\label{fig:photres_CTI1}
\end{figure*}

\begin{figure*}
\includegraphics[width=14cm]{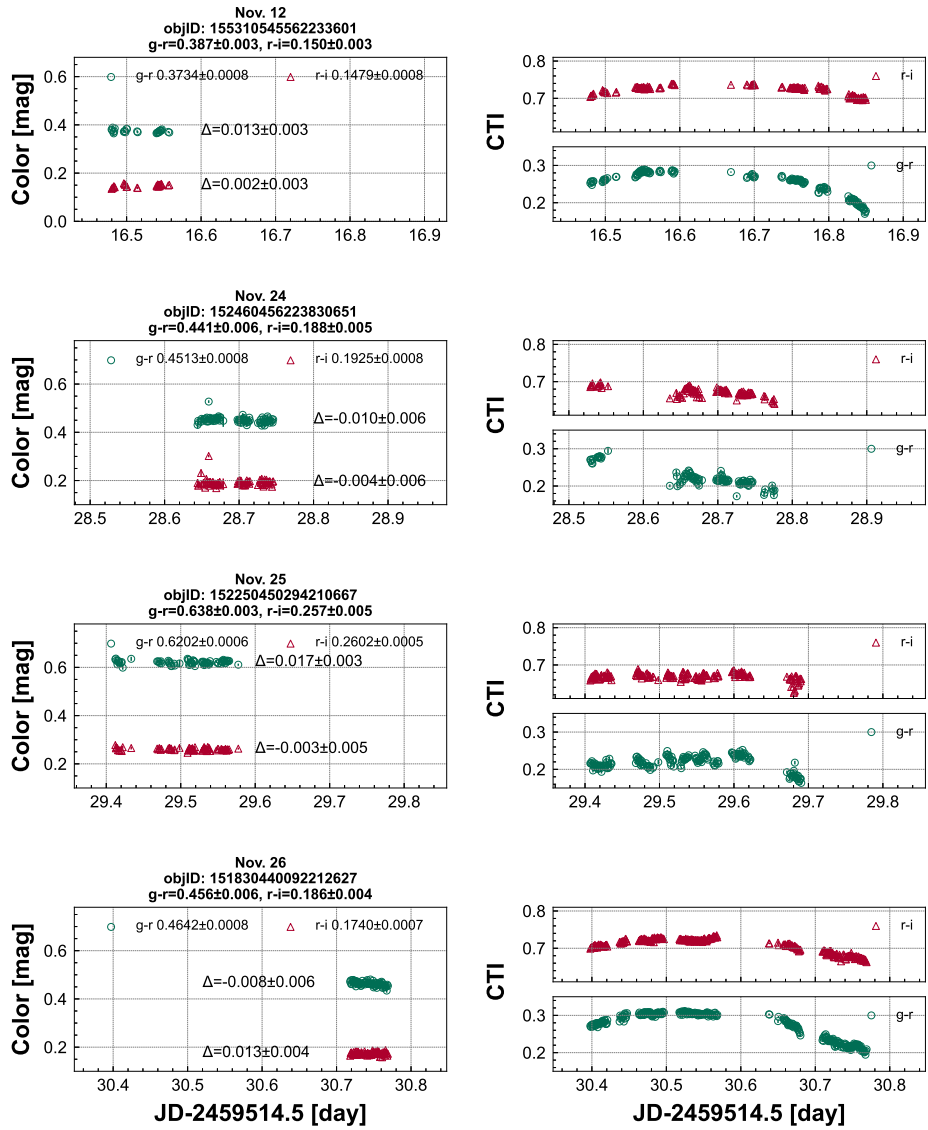}
\caption{%
    Same as figure \ref{fig:photres_CTI1} but for November 12, 24, 25, and 26.}
\label{fig:photres_CTI2}
\end{figure*}

The photometric results of the reference stars and CTIs are presented in the figures 
\ref{fig:photres_CTI1} and \ref{fig:photres_CTI2}.
The color deviations between those in the Pan-STARRS DR2 
and those derived here are at most 0.02 mag for the reference stars on October 28, November 10, 12, 25, and 26.
Thus, we presume the systematic uncertainty of the colors is 0.02 mag in both $g-r$ and $r-i$.
We consider October 28, November 10, 12, 25, and 26 as photometric nights
since CTIs are stable on those days.


%%% Show all authors when n_author <= 8
\newcommand{\noopsort}[1]{} \newcommand{\singleletter}[1]{#1}

\bibliographystyle{pasjbib}

\end{document}